\documentstyle[epsfig,11pt]{article}
\title{\bf Chaotic inflation from a scalar field in non-classical states}
\author{F. Finelli, G. P. Vacca, G. Venturi\\
        Dept. of Physics and INFN, Bologna, Italy.}
\def\beq{\begin{equation}}
\def\eeq{\end{equation}}
\def\bea{\begin{eqnarray}}
\def\eea{\end{eqnarray}}

\def\de{{\rm d}}
\def\De{{\rm D}}
\newcommand{\ket}[1]{\mbox{$\mid #1\,\rangle$}}
\newcommand{\bra}[1]{\mbox{$\langle\, #1 \mid$}}
\newcommand{\expec}[1]{\mbox{$\langle\, #1\,\rangle$}}      
\begin{document}
\maketitle

\abstract{We study chaotic inflation driven by a real, massive, homogeneous 
minimally coupled scalar field in a flat Robertson-Walker spacetime. 
The semiclassical limit for gravity is considered, whereas the scalar field
is treated quantum mechanically by the technique of invariants in order 
to also investigate
the dynamics of the system for non-classical states of the latter.
An inflationary stage is found to be possible for a large set of initial 
quantum states,
obviously including the coherent ones. States associated with a vanishing mean
value of the field (such as the vacuum and the thermal) can also 
lead to inflation, however for such states we cannot make a definitive 
prediction due to the importance of higher order
corrections during inflation. The results for the above coupled 
system are described and their corrections evaluated perturbatively.}

\section{Introduction}

A cosmological exponential expansion was proposed in the 
late 70's and early 80's in 
order to solve the problems of the big bang model \cite{brout,guth}. 
Many proposals have been 
made for the realization of such an expansion, called inflation.
The chaotic model proposed by Linde \cite{linde1} was devoted 
to searching for conditions of a scalar field for which inflation may
occur. No restrictions, such as an initial state of thermal 
equilibrum or a minimum of an effective potential as in
previous models \cite{guth, new}, were assumed. 
Indeed the strength of chaotic models 
is their relative insensitivity to initial conditions\cite{linde2}. 
In this sense they are more 
generic than other models, such as new inflation \cite{new}, which 
are more dependent on initial conditions and/or require fine 
tuning of the parameters which enter in the potential of the 
scalar field. 
However, when chaotic models were first presented, it seemed that 
an implausible initial homogeneity was required in order for 
inflation to start. Therefore much effort was dedicated to showing, by 
numerical and/or analytical estimates, that anisotropies, small and 
large inhomogeneities will in general not suppress the onset 
of inflation (see the review \cite{initial} 
for a detailed discussion of this topic).

In this work, limiting ourselves to the homogeneous case, 
we examine the idea of chaotic inflation 
from a semiclassical point of view, in order to enlarge the 
conditions which allow for inflation to arise from quantum mechanical 
initial states. 
Indeed, most of the chaotic models are also based on a 
classical behaviour for the scalar field.
The "classicality" of the scalar field is a rather questionable assumption:
let us note that the realization of an inflationary domain,
usually as large as several horizon radii \cite{initial}, implies that one is 
considering a region of space-time characterized by a typical size much 
smaller than the Compton wavelength ($1/m$) of the scalar field ($m \ll 
H$ is usually considered in order to obtain sufficiently small 
density perturbations during inflation \cite{rev, brand}). On considering 
gravitational interactions it is difficult to think of the scalar field as 
being in a classical state when gravity effects it on distances smaller 
than its Compton wavelength.

For the above reasons, it is of interest to study the evolution of 
the system for non classical states of the scalar field also.
Here we present a quantum mechanical treatment of the homogeneous 
scalar field which drives a chaotic inflation model with a quadratic 
potential associated with its mass, $V(\phi) = m^2\,\phi^2/2$.
This is greatly facilitated by the possibility of obtaining a Fock 
space of exact 
solutions to the Schr\"odinger equation for the matter field by solving the 
related harmonic oscillator problem with time dependent parameters, 
using the technique of invariants \cite{lewis}. 
Our intention is to investigate whether quantum states of the scalar 
field exist which lead to a sufficient inflationary phase. 

In a certain sense  
we are analyzing "phenomenologically" the inflationary stage 
which is originated by a variety of possible quantum states.
Therefore our conclusions could be of interest also for 
quantum cosmology, whose goal is the initial distribution of 
the scalar field after the quantum era \cite{qc}. 
A classical initial distribution 
was considered in most of the cases treated 
in literature, except for some rare examples where quantum uncertainty
was used to suggest a spread-out distribution for the classical value 
\cite{barvinsky}. 

We shall find an inflationary stage, sufficient to solve the problems of 
standard cosmology, also for states different from 
coherent ones \cite{klauder,coh}.
In fact there is also the possibility that states 
such as the vacuum or a thermal ensemble (which 
have a zero expectation value for the scalar field) 
lead to an inflationary stage for which the time derivative of the
Hubble parameter $H$ is constant and negative.
Indeed the time behaviour for $H$, which satisfies an interesting 
effective equation, is the same as previously found for the 
classical coupled system during inflation 
\cite{piran1}.
However we are not able to study the evolution of the above non-classical 
states for arbitrary times because of quantum corrections 
(due to fluctuations) to the usual equations for matter. 

The above mentioned fluctuations arise naturally from the Born-Oppenheimer 
(BO) reduction \cite{brout2,bv}, which we use to study the 
evolution of the matter-gravity system. In some cases these quantum 
corrections are 
not small and force the scalar field out of the initial quantum state. 
It is not clear whether this effect also alters the above mentioned 
qualitative behaviour of the scale factor during inflation. 

The outline of the paper is as follows: we set up our formalism in 
section 2 which is divided into four paragraphs, of which the first 
two treat the Wheeler-De Witt (WDW) equation \cite{dewitt} for the
Robertson-Walker (RW) minisuperspace considered and its BO reduction.
In the third paragraph 
we describe the method of invariants for time dependent harmonic oscillators 
and in the last we estimate the corrections to the usual matter and gravity 
equations of motions while relegating some useful formulae for their 
computation to the appendix. It is clear that in doing this 
we are taking seriously the WDW 
equation, which represents the full quantum version of this homogeneous 
cosmological toy model. The inclusion of all the Fourier modes 
of the scalar field would certainly represent 
a more reliable version of what is being studied here.
However, we think that consideration of an exact quantum problem 
(although a finite dimensional one rather than an infinite), from 
which the semiclassical approximation can be obtained and verified  
is an important point worth studying.
Section 3 is devoted 
to the study of the dynamical system. In the fourth section we present 
the numerical analysis and summarize our results in the Conclusions.


\section{Basic equations.}
\subsection{The classical system}

We shall study one of the simplest cases of chaotic inflation, which is based 
on the theory of a real free massive scalar field $\phi$ minimally coupled 
to gravity. The action is:

\beq
S \equiv \int d^4x {\cal L} = \int d^4x \sqrt{-g} \left[ \frac{R}{16{\pi}G} 
- \frac{1}{2} g^{\mu \nu} 
\partial_{\mu} \phi \partial_{\nu} \phi
- \frac{1}{2} m^2 \phi ^2 \right] \,
\label{action}
\eeq  
where ${\cal L}$ is the lagrangian density and $m$ is the inverse Compton
wavelength of the field $\phi$.
Further we consider a RW line element with flat spatial section:

\beq
ds^2 = g_{\mu \nu} dx^{\mu} dx^{\nu} = - N^2(t) dt^2 + a^2(t) d\vec{x}^2
\label{metric}
\eeq

\noindent
where $N(t)$ is the lapse function and $a(t)$ is the scale factor.  
On requiring homogeneity for the scalar field, we obtain through an 
Arnowitt-Deser-Misner decomposition,
the following form of the action (\ref{action}):

\beq
S = \int dt d\vec{x} N \, a^3 \left[
\frac{1}{16{\pi}G}
\left( ^{(3)} R + K_{ij} K^{ij} - K^2 \right) 
+ \frac{1}{2} \left( \frac{1}{N^2}  \left( \frac{\partial \phi}
{\partial t} \right)^2 -m^2 \phi^2 \right)
\right] \,.
\label{action2}
\eeq

\noindent
where we have neglected boundary terms.
The {\em three-curvature} scalar $^{(3)} R$, the square of the {\em 
extrinsic curvature tensor} $K_{ij}$ and the {\em extrinsic curvature scalar} 
$K$ for the metric line element (\ref{metric}) are respectively:
                       
\beq
^{(3)} R = 0 \,, \, K_{ij}K^{ij}=\frac{3}{N^2} \, \frac{\dot{a}^2}{a^2} \,, \, 
K=- \frac{3}{N} \frac{\dot{a}}{a} \,,
\label{geom}
\eeq
where a dot denotes the derivative with respect to the time $t$. 
By using the relations 
(\ref{geom}) in the expression for the action (\ref{action2}) we obtain the 
following lagrangian:

\beq
L = N\, V \left[ - \frac{M_{pl}^2}{2} a\, \frac{\dot{a}^2}{N^2} 
+ \frac{a^3}{2} \left( \frac{\dot \phi^2}
{N^2}  -m^2 \phi^2 \right) \right] \,,
\label{lag}
\eeq
where $V$ is the volume of three space, $M_{pl}^2=3/4 \pi G$  and $N$ plays
the role of 
a Lagrange multiplier and is not a dynamical variable.
The Hamiltonian is given by 

\beq
H_{tot} = N\, V \, \left[ - \frac{1}{2 M^2_{pl}} \frac{\pi_a^2}{a} + 
\frac{\pi^2_{\phi}}{2 \, a^3} + a^3 \frac{m^2}{2} \phi^2 \right]
\equiv N\, V \, \left[ - \frac{1}{2 M^2_{pl}} \frac{\pi_a^2}{a} + H_M \right]
\,,
\label{ham}
\eeq 
where $\pi_a = - M_{pl}^2 a \dot{a}/N$ and $\pi_{\phi} = a^3 
\dot{\phi}/N$ are the momenta conjugate to $a$ and $\phi$ 
respectively. The presence of $N$ as a factor in (\ref{ham}) reveals that the 
total Hamiltonian has to be zero: this equation is usually called 
{\em Hamiltonian constraint} 

\beq
\frac{\partial H_{tot}}{\partial N} = 0 \,
\label{ham2}
\eeq
and describes the dynamics,
while the so called {\em momentum constraints} are trivially zero in 
this minisuperspace model. 
In order to quantize the system we rescale $a \to a V^{1/3}$
absorbing the $V$ factor into the Hamiltonian. Henceforth we shall 
work with an $a$ having the canonical dimension of length.

\subsection{The quantum system and its BO reduction}
We start from a fully quantized gravity-matter system 
and subsequently return to a system in which 
gravity is semi-classical \cite{banks} and coupled to quantized matter, as
is illustrated in \cite{brout, bv}. With a coherent state 
\cite{klauder,coh} for matter it is also possible to return to the 
completely classical case.
  
On quantizing the Hamiltonian constraint one obtains 
the WDW equation \cite{dewitt}:  

\beq
\hat{H}_{tot} \Psi(a,\phi) \equiv \left( \frac{\hbar^2}{2 M^2_{pl}}
\frac{\partial^2}{\partial a^2}\frac{1}{a} + \hat{H}_M \right)
\Psi(a,\phi) =0
\label{wdw}
\eeq
\noindent
where $\Psi$ is a function of $a$ and $\phi$ and describes both gravity
and matter. Subsequently one performs a BO decomposition of $\Psi$ as:

\beq
\Psi(a, \phi) = a \psi(a) \chi(a,\phi)
\label{bo}
\eeq
where $\chi(a,\phi)$ is not further separable. 
Let us add a brief comment on the quantization of the Hamiltonian constraint 
and on the assumption (\ref{bo}): in order to obtain 
eq. (\ref{wdw}) from eq. (\ref{ham2}) a particular choice 
of factor ordering has been made ($\frac{\pi_a^2}{a} \rightarrow 
\hat{\pi}^2_a \frac{1}{a}$) \cite{cv}. The ordering ambiguity in 
this context has been previously noted \cite{qc}. We observe that our 
choice is convenient since it is associated with eq. (\ref{bo}), which
agrees with 
$\Psi = 0$ for $a \le 0$ as expected. All orderings will coincide in the 
classical limit and differ by ${\cal O} (\hbar)$. We have verified that such 
differences are negligible 
during the inflationary regime and in the semiclassical limit for gravity, 
which we consider.  

An effective equation 
of motion for $\psi$ may now be obtained \cite{bv} 
by first substituting the above
decomposition into eq. (\ref{wdw}), contracting with $\chi^*$ 
and dividing by $\langle \chi | \chi \rangle$. One then has :

\begin{eqnarray} 
\lefteqn{
\left[ \frac{\hbar^2}{2M^2_{pl}} \De^2 + a \frac
{\langle \chi | \hat{H}_M | \chi \rangle}{\langle \chi | \chi \rangle}
\right] \psi = - \frac{\hbar^2}{2 M^2_{pl}}
\frac{\langle \chi | \bar{\De}^2 | \chi \rangle}{\langle \chi | \chi \rangle}
\psi } \nonumber \\ & &
= \frac{\hbar^2}{2 M^2_{pl}} \frac{1}{\langle \chi | \chi \rangle}
\left\{ \langle \chi |
\frac{\stackrel{\leftarrow}{\partial}}{\partial a}
\left( 1 - \frac{| \chi \rangle \langle \chi |}
{\langle \chi |\chi \rangle } \right)
\frac{\partial}{\partial a} |
\chi \rangle \right\} \psi
\label{eqpsi}
\end{eqnarray}
where we have introduced covariant derivatives:

\beq
\De \equiv \frac{\partial}{\partial a} + i {\cal A} \,; \;\;\;\;\;
\bar{\De} \equiv \frac{\partial}{\partial a} - i {\cal A}
\label{code}
\eeq
with:

\beq
{\cal A} \equiv - i \frac{\langle \chi | \frac{\partial}{\partial a} \chi \rangle}
{\langle \chi | \chi \rangle} \equiv - i \langle \frac{\partial}{\partial a}
\rangle
\eeq
and a scalar product:

\beq
{\langle \chi | \chi \rangle} \equiv \int \de \phi \chi^*(a,\phi) \chi(a,\phi)
\eeq
where the integral is over the different matter modes.

An equation for $\chi$ may be obtained \cite{bv} 
by multiplying (\ref{eqpsi}) by $\chi$ 
and subtracting it from (\ref{wdw}):

\begin{eqnarray} \lefteqn{
a \,\psi \left( \hat{H}_M - \langle \hat{H}_M \rangle \right) \chi +
\frac{\hbar^2}{M^2_{pl}} \left( \De \psi \right) \bar{\De} \chi =
- \frac{\hbar^2}{2 M^2_{pl}} \psi \left( \bar{\De}^2 - \langle \bar{\De}^2 
\rangle \right) \chi =
 } \nonumber \\ & &
- \frac{\hbar^2}{2 M^2_{pl}} \psi \left[ \left(
\frac{\partial^2}{\partial a^2} - \langle \frac{\partial^2}{\partial a^2}
\rangle \right) - 2 \langle \frac{\partial}{\partial a} \rangle
\left(
\frac{\partial}{\partial a} - \langle \frac{\partial}{\partial a}
\rangle \right)  \right] \chi \,.
\label{eqchi}
\end{eqnarray}
We note that the r.h.s. of eqs. (\ref{eqpsi}) and (\ref{eqchi}) are
related to fluctuations, that is they consist of an operator acting on
a state minus its expectation value with respect to that state.
Henceforth we shall refer to them as RHS fluctuations.

It is convenient 
to write the above equations in terms of normal rather than 
covariant ($D$) derivatives and take into account 
the contribution from the connections by multiplying the gravity 
and matter wave functions by a phase. This redefinition leaves invariant 
the original wave-function $\Psi$ in (\ref{bo}) since the 
two phases are opposite in sign. 
With 

\beq
\tilde{\psi} \equiv e^{+i \int^a {\cal A} \de a'} \psi
\label{psitil}
\eeq
eq. (\ref{eqpsi}) becomes 

\beq 
\left( \frac{\hbar^2}{2 M^2_{pl}} \frac{\partial^2}{\partial a^2} +
a \, \langle \hat{H}_M \rangle \right) \tilde{\psi} =
- \frac{\hbar^2}{2 M^2_{pl}} \langle \bar{\De}^2 \rangle \tilde{\psi}
\label{eqpsitilde}
\eeq
Further we observe that in the semiclassical (WKB) limit for $\tilde{\psi}$
one has:

\beq
\tilde{\psi} = 
W \, e^{ \frac{i}{\hbar} S_{eff} } \,,
\label{wkb}
\eeq
where $S_{eff}$ is solution to the Hamilton-Jacobi equation 

\beq
- \frac{1}{2 M^2_{pl}} \left( \frac{\partial S_{eff}}{\partial a} \right)^2
+ a \langle \hat{H}_M \rangle = 0 \,,
\label{seff}
\eeq
and $W=(\partial S_{eff}/\partial a)^{-\frac{1}{2}}$. 
We analogously define $\tilde \chi$ by
\beq
\tilde{\chi} \equiv e^{-i \int^a {\cal A} \de a'} \chi
\label{chitil}
\eeq
so that eq. (\ref{eqchi}) can be written in the form
\beq
a \, \tilde{\psi} \left( \hat{H}_M - \langle \hat{H}_M \rangle \right) 
\tilde{\chi} +
\frac{\hbar^2}{M^2_{pl}} \left( \frac{\partial}{\partial a} \tilde{\psi} 
\right) \frac{\partial}{\partial a} \tilde{\chi} =
- \frac{\hbar^2}{2 M^2_{pl}} \tilde{\psi} 
\left( \bar{\De}^2 - \langle \bar{\De}^2 \rangle \right) \chi.
\eeq
On introducing a time derivative 

\beq
i \hbar
\frac{\partial}{\partial t} \equiv - i \frac{\hbar}{M^2_{pl}} \frac{\partial 
S_{eff}}
{a \partial a} \frac{\partial}{\partial a}
\label{time}
\eeq
and defining

\beq
\chi_s \equiv e^{- \frac{i}{\hbar} \int^t \langle \hat{H}_M \rangle dt'
- i \int^a da' {\cal A}} \chi =
e^{- \frac{i}{\hbar} \int^t \langle \hat{H}_M \rangle dt'} \tilde{\chi}
\label{chisch}
\eeq
eq. (\ref{eqchi}) may be further rewritten as: 

\beq
\left( \hat{H}_M - i \hbar \frac{\partial}{\partial t} \right)
\chi_s =
- \frac{\hbar^2}{M^2_{pl} a } e^{ - \frac{i}{\hbar} \int^t \langle 
\hat{H}_M \rangle \de t' -
i \int^a \de a' {\cal A}} \left[ \frac{\partial \log W}{\partial a}
\bar{\De} + \frac{1}{2} \left( \bar{\De}^2 - \langle \bar{\De}^2 \rangle 
\right) \right] \chi \,.
\label{sch1}
\eeq
which is defined where $\tilde \psi$ has support in the semiclassical limit
(eq. (\ref{wkb})) and henceforth we shall indicate by $a$ the classical scale
factor $a(t)$.
We note that on neglecting the rhs of eq. (\ref{sch1}) one has a time
dependent Schr\"odinger equation for $\chi_s$.
It is important to remember that with 
this approximation one does not take in account the transitions of the 
quantum state of the Schr\"odinger equation caused by the RHS.
We shall discuss these transitions in sections 2.4 and 4.

Let us stress that the peculiarities of this coupled system originated from 
the BO reduction of the WDW equation. 
The Schr\"odinger equation, involving time defined in (\ref{time}), describes 
the quantum evolution of matter, which in turn leads to the back-reaction 
appearing in the semiclassical evolution of the scale factor in eq. (\ref{seff}). 
\subsection{The quantum matter}

The quantum evolution of matter, as described by eq. (\ref{sch1}), is 
given by a Hamiltonian:

\beq
\hat{H}_M = - \frac{\hbar^2}{2 a^3} \,
\frac{\partial^2}{\partial \phi^2} + a^3 \frac{m^2}{2} \hat{\phi}^2 \,
\label{matham}
\eeq 
which is  a harmonic oscillator with a time dependent mass $a^3(t)$ and a
constant frequency $m$ and has been previously studied 
(see for instance \cite{lewis,cinesi}) without resorting to the adiabatic
approximation \cite{berry}. 

However, since we are interested in a problem for which the evolution of 
$a(t)$ could have an exponential behaviour, it is much better to employ  
a technique whereby a time-dependent problem can be 
solved {\em exactly} (we work in the Schr\"odinger 
representation). This technique is based on an auxiliary 
{\em invariant} operator (historically called adiabatic invariant
\cite{lewis}) which satisfies 

\beq
i \, \hbar \, \frac{\partial \hat{I}}{\partial t} 
+ [\hat I, \hat H_M]=0 \,.
\eeq 
\noindent
The quadratic hermitean invariant originally
introduced in \cite{lewis} has real time-independent eigenvalues and 
is given by:

\beq
\hat I =
\frac{1}{2} \left[ \frac{\hat \phi^2}{\rho^2} 
+ (\rho \hat \pi_{\phi} - a^3
\dot{\rho} \hat \phi)^2 \right]
\ ,
\label{I}
\eeq
where $\rho$ satisfies the following equation:

\beq
\ddot{\rho} + 3 \frac{\dot a}{a} \dot{\rho}
+ m^2 \rho = \frac{1}{a^6 \rho^3}
\label{rho}
\eeq
This invariant $\hat{I}$ is important and useful
since a general solution of the time dependent Schr\"odinger 
equation with the Hamiltonian (\ref{matham}) can be expressed as:
\beq
\ket{\chi,t}_s = \sum\limits_n c_n\,
e^{i\,\varphi_n(t)}\,\ket{n, t}
\ ,
\label{general}
\eeq
where the $c_n$ are time-independent coefficients, $\ket{n, t}$ are the 
eigenstates of the invariant (\ref{I}) with real eigenvalues $\lambda_n$ 
and the phases $\varphi_n(t)$ are given by 
\beq
\varphi_n(t)=\frac{1}{\hbar} \int_{t_0}^t 
\bra{n,t'}\,i\hbar\,\partial_{t'}-\hat H_M(t')\,\ket{n,t'}
\, dt'
\ .
\label{fase}
\eeq

The quadratic hermitean invariant (\ref{I}) can be decomposed in basic linear 
non-hermitean invariants \cite{cinesi}:

\beq
\begin{array}{c}
\hat I_b (t) = 
e^{i\Theta (t)} \hat b (t) = { e^{i\Theta} \over \sqrt{2\,\hbar}}\,
\left[{\hat \phi \over \rho} +
i\,\left(\rho\,\hat \pi_{\phi} - a^3 \dot{\rho} \hat \phi \right)\right] \,,
\\
\\
\hat I_b^{\dagger} (t) =
e^{-i\Theta (t)} \hat b^{\dagger} (t) = { e^{-i\Theta} \over
\sqrt{2\,\hbar}}\,\left[{\hat \phi \over \rho}-
i\,\left(\rho \,\hat \pi_{\phi} - a^3 \dot{\rho} \hat \phi \right)\right]
\,,
\end{array}
\eeq
with 
\beq
\Theta (t)=\int_{t_0}^t
{dt'\over a^3(t')\,\rho^2(t')} \,.
\eeq
The quadratic invariant (\ref{I}) can then be written as:

\beq 
\hat I (t) =
\hbar \left(\hat I_b^{\dagger} (t)\,\hat I_b (t) + \frac{1}{2} \right) = 
\hbar \left(\hat b^{\dagger} (t) \,\hat b (t) + \frac{1}{2} \right)
\ ;
\eeq
with $[\hat I_b,\hat I_b^{\dagger}]=
[\hat b,\hat b^{\dagger}]=1$ which allows us to identify the 
pairs of creation and destruction operators, $(\hat I_b, 
\hat I_b^{\dagger})$ and $(\hat b,\hat b^{\dagger})$. The first is 
used to define a Fock space describing the solutions 
to the Schr\"odinger equation, and the second one  
generates the space of the eigenstates of the invariant defined in (\ref{I}). 
As already shown in \cite{bertoni} this difference is important because 
$\chi$, as defined in (\ref{bo}), belongs to the second space, whereas 
$\chi_s$, as defined in (\ref{chisch}), belongs to the first, 
of course always on neglecting the RHS fluctuations.  

Let us emphasize that the $\hat{b}$ operators that factorize the quadratic 
invariant are quite different from the creation and distruction operators
(which we may denote by $\hat{d}$) that factorize the Hamiltonian 
(\ref{matham}):

\beq
\hat{H}_M = \hbar m \left(\, \hat d^{\dagger} (t) \, \hat d(t)
+ \frac{1}{2}\,\right)
\eeq
with:
\beq
\begin{array}{c}
\hat{d} = \left( \frac{m a^3}{2 \hbar} \right)^{\frac{1}{2}}
\left( \hat\phi + i\,\frac{\hat\pi_{\phi}}{m a^3} \right) \,,
\\
\\
\hat d^{\dagger} = \left( \frac{m a^3}{2 \hbar} \right)^{\frac{1}{2}}
\left( \hat \phi - i\,\frac{\hat \pi_{\phi}}{m a^3} \right)
\ .
\end{array}
\eeq
and $[\hat d, \hat d^{\dagger}]=1$. The two sets of $d$ and $b$ operators are 
related by the following Bogoliubov transformation

\begin{eqnarray}
\hat d &=& 
B^* (t)\, \hat b + A^* (t)\,\hat b^{\dagger} 
\equiv \nonumber \\
&\equiv& \frac{1}{2} \hat b \left[ \rho (a^3 m)^{\frac{1}{2}} +
\frac{1}{ \rho (a^3 m)^{\frac{1}{2}}} + i \dot{\rho} \left(\frac{a^3}
{m}\right)
^{\frac{1}{2}} \right] + \nonumber \\
&+& \frac{1}{2} \hat b^{\dagger} 
\left[ \rho (a^3 m)^{\frac{1}{2}} - \frac{1}{\rho (a^3 m)^
{\frac{1}{2}}}
+ i \dot{\rho} \left(\frac{a^3}{m}\right)
^{\frac{1}{2}} \right]  \,,
\label{bogo}
\end{eqnarray}
and similarly for $\hat d^{\dagger}$.

Since the quantum mechanical problem related to this time-dependent 
harmonic oscillator can be solved exactly, as in eq. (\ref{general}), 
let us first consider the state which best represents the classical 
behaviour of the scalar field in order to reproduce the 
previous results \cite{madsen, russi, piran1, amend}. As is known 
\cite{coh} this state is a {\em coherent} state \cite{klauder} for $\hat I_b$, 
which satisfies:

\beq
\hat I_b \ket{\alpha, t}_s = \alpha \ket{\alpha, t}_s
\eeq  
where $\alpha \equiv |\alpha|\,e^{i\beta}$ is a time-independent 
complex c-number. 
Such a state satisfies the Schr\"odinger equation and can also be written 
as a normalized superposition of eigenstates of the quadratic invariant in 
(\ref{I}):

\beq
\ket{\alpha,t}_s = e^{-|\alpha|^2/2}\,
\sum\limits_{n=0}^\infty\,{\alpha^n\over\sqrt{n!}}\,
\ket{n,t}_s =
e^{-|\alpha|^2/2 - i\Theta/2}\,
\sum\limits_{n=0}^\infty\,{\alpha^n\,e^{-i\,n\,\Theta}\over\sqrt{n!}}\,
\ket{n,t}_b
\ .
\label{coherent}
\eeq
On defining $\expec{\hat O}_{\alpha} \equiv \, _s \bra{\alpha,t}\,\hat O \,
\ket{\alpha,t}_s$, the mean value of some physically relevant 
quantities is:

\beq
\phi_c \equiv \expec{\hat \phi}_{\alpha} =
\sqrt{2\,\hbar}\,|\alpha|\,\rho\,\cos(\Theta-\beta)
\ ,
\label{ampl}
\eeq

\beq
\pi_{\phi,c} \equiv
\expec{\hat\pi_\phi}_{\alpha} =
\sqrt{2\,\hbar}\,|\alpha|\,\left[
a^3\,\dot \rho \,\cos(\Theta-\beta)
-{1\over \rho}\,\sin(\Theta-\beta)\right]
\,
\eeq

\beq
\begin{array}{c}
\expec{(\Delta\hat\phi)^2 }_{\alpha} \equiv
\expec{(\hat \phi-\phi_c)^2}_{\alpha}
=\strut\displaystyle{\hbar\over 2}\,\rho^2 \,  \\
 \\
\expec{(\Delta\hat\pi_\phi)^2}_{\alpha}\equiv
\expec{(\hat \pi_\phi-\pi_{\phi,c})^2}_{\alpha}
=\strut\displaystyle{\hbar\over 2}\,
\strut\displaystyle\left({1\over \rho^2}+a^6\,\dot \rho^2\right)
\ , \\
\\
\expec{(\Delta\hat\phi)^2 }_{\alpha}^{\frac{1}{2}}
\,\expec{(\Delta\hat\pi_\phi)^2}_{\alpha}^{\frac{1}{2}} =
{\hbar\over 2}\,\sqrt{1+a^6\,\rho^2\,\dot \rho^2}
\ .
\label{indet}
\end{array}
\eeq
We note that the mean value of the field (\ref{ampl}) satisfies 
the classical equation of motion:
\beq
\ddot \phi_c+3\,{\dot a\over a}\,\dot \phi_c
+m^2\,\phi_c=0
\ ,
\eeq
and $\alpha$ does not appear in the uncertainty relation (\ref{indet}), 
which however increases with time. 

The mean value of the matter Hamiltonian for the coherent state is:

\begin{eqnarray}
\expec{\hat H_M}_{\alpha} &=&
\frac{1}{2\,a^3}\,\left[ 
\expec{\hat \pi_\phi}_{\alpha}^2+
m^2 a^6 \, \expec{\hat \phi}_{\alpha}^2 + 
\frac{\hbar}{2}
\left( \dot \rho^2 \, a^6 + \frac{1}{\rho^2} + m^2 \, \rho^2 \, a^6\right)
\right] \nonumber \\&=&
H_{cl} + H_0 \equiv a^3 (\mu_{cl} + \mu_0) \,
\label{cohehami}
\end{eqnarray}
where $H_{cl}$ is the classical Hamiltonian for the scalar field 
$\phi_c$ defined in eq.(\ref{ampl}), $H_0$ is the zero-point Hamiltonian 
and $\mu_{cl}$, $\mu_0$ are respectively the classical and zero point 
energy-densities. All these quantities are time-dependent.
In order to consider the above semiclassical field as a fluid we further 
define, in analogy to the classical case, a {\em pressure} operator:

\beq 
\hat p = \frac{1}{2\,a^3}\left( \frac{\hat{\pi}^2_{\phi}}{a^3} - 
m^2 a^3 \hat \phi^2 \right)\,.
\label{qpress}
\eeq
Its expectation value in a coherent state is: 

\beq
\expec{\hat p}_{\alpha} = \frac{1}{2\,a^3} \left[ \frac{ \pi^2_ {\phi,c} }
{a^3} - m^2 a^3 \phi^2_c + \frac{\hbar}{2} 
\left( \dot \rho^2 \, a^3 + \frac{1}{a^3 \rho^2} - m^2 \, \rho^2 
\, a^3 \right) \right] 
\equiv p_{cl} + p_0 \,
\label{press}
\eeq
and we have the usual fluid conservation law:

\beq
\dot \mu_{cl} + \dot \mu_0 + 3 \frac{\dot a}{a} 
\left(\mu_{cl} + \mu_0 + p_{cl} + p_0 
\right) = 0 \,.
\eeq

\subsection{Higher order corrections to semiclassical theory}

We are interested in the corrections given by the rhs of 
eq. (\ref{eqpsi}) for gravity and by the rhs of eq. (\ref{eqchi}) 
for $\chi$ in a coherent state or an n-particle state \cite{cooper},
for which $\langle \chi | \chi \rangle = 1$. 
For this purpose we note that the $\chi$ wave function introduced in 
(\ref{bo}) can be constructed by using the $b,\, b^{\dagger}$ operators. 
Let us then indicate by $\ket{\alpha}_b$ the coherent state associated 
with the operator $\hat{b}$ while the $\ket{n}_b$ are the eigenstates of the 
associated number operator.
Further, we shall derive expressions valid during the inflationary regime
($a >> 1$), so that some simplification will occur.

Let us first consider the corrections to the gravity equation.
With the introduction of the semiclassical time in eq. (\ref{time}), 
we can write:

\beq 
\bra{\chi} \bar{\De^2} \ket{\chi} = \frac{1}{\dot a^2} \left( 
|\frac{\partial}{\partial t} \ket{\chi}_b |^2 - |_b\bra{\chi}
\frac{\partial}{\partial t} \ket{\chi}_b |^2 \right)=
\frac{1}{\dot{a}^2}P_\chi \,.
\eeq
On defining the Hubble parameter as $H= \frac{\dot a}{a}$, 
the semiclassical version of eq. (\ref{eqpsitilde}) leads to:
\beq
H^2 = 
\frac{8 \pi G}{3 a^3} \bra{\chi} \hat{H}_M \ket{\chi} +
\frac{\hbar^2}{a^4 M^4_{pl}} \bra{\chi} \bar{\De}^2 \ket{\chi} 
=\frac{8 \pi G}{3 a^3} \langle \hat{H}_M \rangle +
\frac{\hbar^2}{a^6 M^4_{pl} H^2} P_\chi \,.
\label{semcorr}
\eeq
With the help of formulae in the appendix one obtains:
\beq
P_n=|\frac{\partial}{\partial t} \ket{n}_b |^2 - | _b\bra{n}
\frac{\partial}{\partial t} \ket{n}_b |^2 = 
2 m^2 |A|^2 |B|^2 (n^2+n+1)
\label{gravn}
\eeq
\bea 
P_\alpha=|\frac{\partial}{\partial t} \ket{\alpha}_b |^2 - | _b\bra{\alpha}
\frac{\partial}{\partial t} \ket{\alpha}_b |^2 &=& \left[ 
\dot\Theta - m \left(1 + 2|A|^2\right) \right]^2 \, |\alpha|^2 
+ 2 m^2 |A|^2 |B|^2 
\left(1 + 2 |\alpha|^2\right) + \nonumber \\
&+& 2 m \left( \alpha^2 A B^* 
+ \alpha^{* \, 2} B A^* \right) 
\left( m \left(1 + 2|A|^2\right) - \dot\Theta \right) \,.
\eea  
We note that for the coherent case the terms in $\alpha$ of 
order higher than quadratic cancel.
For completeness we also write the average matter Hamiltonian for an 
$\ket{n}_b$ state:
\beq
_b\bra{n} \hat{H}_M \ket{n}_b=\hbar m (|A|^2+\frac{1}{2})(2n+1)=
a^3 \mu_0 (2n+1)
\label{n}
\eeq
and on solving eq. (\ref{semcorr}) for $H^2$ one obtains 
\beq
H^2=\frac{2 \langle \hat{H}_M \rangle}{M^2_{pl} a^3}
\Biggl( \frac{1+\sqrt{1+\frac{\hbar^2 P_\chi}{\langle \hat{H}_M \rangle^2}}}{2}
\Biggr)=\frac{2 \langle \hat{H}_M \rangle}{M^2_{pl} a^3} \xi
\eeq 

Let us now estimate the corrections, for $a \gg 1$, 
to the results obtained when the rhs is neglected
for the coherent, the $\ket{n}_b$ and the vacuum states respectively. 
One first observes that the time dependence of the rhs in 
this limit is the same as that of the Hamiltonian, so that their ratio
becomes time independent. 
For the coherent case in the $|\alpha| \to \infty$ limit we have:
\beq
\frac{\hbar^2 P_\alpha}{_b\bra{\alpha}\hat{H}_M\ket{\alpha}_b^2}  
\approx
 \frac{2+\cos{2\beta}}{4 |\alpha|^2}
\eeq
which goes to zero, so that one may expect no correction in the classical 
limit.
For the $\ket{n}_b$ state case, in the $n \to \infty$ limit, we have
\beq
\frac{\hbar^2 P_n}{_b\bra{n}\hat{H}_M\ket{n}_b^2}  \approx \frac{1}{2}
\eeq
which leads to an $11\%$ correction. 
Finally for the vacuum case ($\alpha=n=0$) we obtain:
\beq
\frac{\hbar^2 P_0}{_b\bra{0}\hat{H}_M\ket{0}_b^2}  \approx 2 
\eeq
which gives a correction $(\xi-1)$ of order $36\%$.

We now consider the matter equation.
It is convenient to introduce
\beq
\ket{\phi_{\chi_{2},\chi_{1}}}=\ket{\chi_2} - \bra{\chi_1} \chi_2 \rangle 
\ket{\chi_1} .
\eeq
As the corrections are an operator we shall examine their matrix 
elements with respect to the $\ket{n}_b$ states and
the coherent states (an overcomplete non-orthogonal set). 
The general form will be
\bea
\frac{\hbar^2}{2M^2_{pl}a\dot{a}^2} \Bigl[ -2 
\frac{\partial W}{\partial t} \bra{\phi_{\chi_2,\chi}} 
\frac{\partial}{\partial t} \ket{\chi} -
\dot{a}^2 \frac{\partial}{\partial a}(\bra{\phi_{\chi_{2},\chi}} 
\frac{\partial}{\partial a} \ket{\chi})+ \nonumber \\
\bra{\phi_{\chi_{2},\chi}} \mathop{\frac{\partial}{\partial t}}^{\leftarrow}
\frac{\partial}{\partial t} \ket{\chi} +
2 \bra{\chi} 
\frac{\partial}{\partial t} \ket{\chi} \bra{\phi_{\chi_2,\chi}} 
\frac{\partial}{\partial t} \ket{\chi} \Bigr]
\label{matcorr}
\eea
\noindent
In the large $a$ limit, during inflation,
the first two terms of eq. (\ref{matcorr}) will be subdominant
with respect the last two by a factor of order $a^3$, hence we may neglect 
them and denote the last two by $Q_{\chi_2,\chi}$.

Using the results given in the appendix and in eqs. (\ref{n}) and
(\ref{matcorr}) one has for the ratio of the
the matrix elements between the $\ket{n}_b$ and $\ket{m}_b$ states 
and the average Hamiltonian :
\bea
\frac{Q_{m,n}}{_b\bra{n} \hat{H}_M \ket{n}_b} = \frac{1}{4 \xi (2n+1)^2} \Bigl[
-4\sqrt{n(n-1)}\delta_{m,n-2}+\sqrt{n(n-1)(n-2)(n-3)}\delta_{m,n-4}+
\nonumber \\
\sqrt{(n+1)(n+2)(n+3)(n+4)}\delta_{m,n+4}+ 4 \sqrt{(n+1)(n+2)}\delta_{m,n+2}
\Bigr]\,
\label{matrixn}
\eea
which are of order unity for the vacuum ($n=0$) state and one order 
smaller, but finite, for large $n$. 
The form obtained reveals that the allowed transitions 
will eventually lead the system to spread over a superposition of states

In the same manner we obtain expressions for the coherent states in 
the large $a$ limit. Only the last two terms of eq. (\ref{matcorr})
contribute giving
\bea
Q_{\beta,\alpha}&=&\frac{_b\bra{\beta} \alpha \rangle_b \hbar^2 m^2 |A|^4}
{2 M^2_{pl} a^3 H^2} (\beta^*-\alpha^*)
\Bigl[
(\alpha^*+\beta^*+4\alpha)(4+\beta^{*2})- \nonumber \\
&&\alpha^*(\alpha^{*2}+\beta^{*2})
+\alpha^2(8\alpha+11\alpha^*+3\beta^*)+4\alpha \alpha^{*2}
\Bigr]
\eea
In order to see how large such corrections are in the 
$|\alpha| \rightarrow \infty$ 
limit, when the system is characterized by a behaviour close to the 
classical, we set $\gamma \equiv \beta - \alpha$ and obtain 
\bea
\frac{Q_{\beta,\alpha}}{_b\bra{\alpha}\hat{H}_M \ket{\alpha}_b}&=&
\frac{e^{-|\gamma|^2/2+i \, Im(\alpha \beta^*)} \hbar^2 m^2 |A|^4}
{4 \xi \, _b\bra{\alpha}\hat{H}_M \ket{\alpha}_b^2} \gamma^*
\Bigl[
8\alpha^* +16\alpha+4\gamma^*+3(\alpha^2+\alpha^{*2}+\gamma^*\alpha^*
+ \nonumber \\
&& \gamma^{*3}+ 
 8\alpha\alpha^{*2}+8|\alpha|^2\gamma^*+4\alpha\gamma^{*2}+
8\alpha^3+14\alpha^*\alpha^2 \Bigr] 
\mathop{\sim}_{|\alpha|>>1} {\cal O}(\frac{1}{|\alpha|}).
\eea
We note that the above quantity is bounded in $\gamma$ so that in the
$|\alpha| \to \infty$ limit every transition away from a coherent state is
suppressed and thus, as expected, one recovers the classical behaviour 
of the system. Further the behaviour ${\cal O}(\frac{1}{|\alpha|})$ is 
also valid for any $a$.

Let us finally exhibit the ratio for the vacuum case ($\alpha=0$):
\beq
\frac{Q_{\beta,0}}{_b\bra{0}\hat{H}_M \ket{0}_b}=
\frac{_b\bra{\beta}0\rangle_b \hbar^2 m^2 |A|^4}
{4 \xi _b\bra{0}H_M\ket{0}_b^2}(\beta^{*4}+4\beta^{*2})=
\frac{1}{4 \xi} (\beta^{*4}+4\beta^{*2}) e^{-|\beta|^2/2}.
\eeq
Of course if one multiplies the above by 
$_b\bra{n}\beta\rangle_b$ and integrates 
$\int d\mu(\beta)$, where the measure ($d\mu(\beta)
=\frac{d{\cal R}e \beta \, d{\cal I}m \beta}{\pi}$) 
is such that the completeness relation is 
$\int d\mu(\beta) \ket{\beta}_b \, _b\bra{\beta} =1$, we re-obtain the matrix
elements for $n=0$ in eq. (\ref{matrixn}) as expected.

Since we are also interested in the time evolution for the system initially
in the vacuum state we shall also estimate in section $4$ the transition rates 
(to the $\ket{2}_b$ and $\ket{4}_b$ states) 
which are generated by the correction terms on the rhs of the matter
equation.

\section{The analysis of the dynamical systems}

In this section, in preparation for the numerical analysis,
we first derive an effective equation describing the evolution of 
semiclassical gravity after ``averaging'' over the quantum matter degree
of freedom. In our model, on neglecting the RHS fluctuations, this equation 
will not depend on the matter quantum state. 
Subsequently we shall study the dynamics of the matter quantum state
in order to understand and possibly evaluate the effect of the 
approximations made. In particular we shall consider in detail the evolution 
for a coherent state, a vacuum state and a thermal state.

\subsection{ Effective equation for gravitation.}

Let us begin by considering matter in a generic quantum state 
$\ket{\chi}_s$ satisfying the Schr\"odinger equation with Hamiltonian 
$\hat{H}_M$. The backreaction on gravity is described by the equation
\beq
H^2=
\frac{8 \pi G}{3 a^3} {_s\bra{\chi}\hat{H}_M\ket{\chi}_s}=
\frac{8 \pi G}{3}{_s\bra{\chi}\hat{\mu}\ket{\chi}_s} \equiv \frac{8 \pi G}{3} 
\mu
\label{hub2} 
\eeq
where $\hat{\mu}=\hat{H}_M/a^3$ is the energy density operator
and $\mu$ its average.
On differentiating equation (\ref{hub2}) with respect to time one obtains:
\beq
\dot{H}+\frac{3}{2}H^2=\frac{4 \pi G}{3 a^3}
{_s\bra{\chi}\frac{1}{H}\frac{\partial \hat{H}_M}{\partial t} \ket{\chi}_s}=
- 4 \pi G {_s\bra{\chi}\hat{p}\ket{\chi}_s} \equiv 
- 4 \pi G p
\label{passo1}
\eeq
where $\hat{p}$ is the pressure operator defined in (\ref{qpress}) and
$p$ its average.
One notes that $\dot{H}=- 4 \pi G (p+\mu)$ so that a fluid-like equation is
satisfied:
\beq
\dot{\mu}+3 H (p+\mu)=0
\eeq

We now differentiate eq. ({\ref{passo1}) with respect to time and use the 
relations (\ref{hub2}) and
(\ref{passo1}) in order to eliminate the average, with respect to 
$\ket{\chi}_s$, of the Hamiltonian and of its time derivative. One then obtains
\beq
\ddot{H}+6 H \dot{H}=\frac{4 \pi G}{a^3} m^2
{_s\bra{\chi}\{ \hat{\phi},\hat{\pi}_{\phi} \}\ket{\chi}_s}
\label{passo2}
\eeq
where the term on the RHS comes from the commutator $[ \hat{p},\hat{H}_M ]$.
Further we perform yet another time derivative of (\ref{passo2})
obtaining
\beq
\mathop{H}^{\dots} + 6 \dot{H}^2 + 6 H \ddot{H}=
-\frac{12 \pi G}{a^3}m^2 H
{_s\bra{\chi}\{ \hat{\phi},\hat{\pi}_{\phi} \}\ket{\chi}_s}
+ \frac{4 i \pi G}{ \hbar a^3}m^2
{_s\bra{\chi}[\hat{H}_M,\{ \hat{\phi},\hat{\pi}_{\phi} \} ]\ket{\chi}_s}
\label{passo3}
\eeq
One easily computes the commutator in the last term
$[\hat{H}_M,\{ \hat{\phi},\hat{\pi}_{\phi} \} ] = 
-4 i \hbar a^3 \hat{p}$, then,
on using the relations (\ref{passo1}) and (\ref{passo2}), one finally 
obtains an equation for $H$ only:
\beq
\mathop{H}^{\dots} +6 \dot{H}^2 +4 \dot{H} + 9 \ddot{H} H + 18 H^2 
(\dot{H} +\frac{1}{3}) = 0 \,,
\label{order3}
\eeq
where we have scaled the time by $m^{-1}$ and 
henceforth we shall use this new scaled time.

The above equation is valid for every quantum state, 
in particular for a coherent state, and, 
on taking the classical limit for matter, also for 
the classical field. Thus the classical model of chaotic inflation can be
described by this equation.
Indeed, it is known that the classical system  is governed by the 
system of equations
\bea
H^2=
\frac{8 \pi G}{3}\Bigl( \frac{\phi_c^2+\dot{\phi_c}^2}{2}\Bigr) \nonumber \\
\ddot{\phi_c}+3\dot{\phi_c}
\sqrt{\frac{8 \pi G}{3}\Bigl( \frac{\phi_c^2+\dot{\phi_c}^2}{2}\Bigr)} +
\phi_c=0
\label{classic}
\eea
which gives a non linear second order differential equation 
in $H$ but, as is easy to check, also implies eq. (\ref{order3}).
Hence the classical system has a dynamics which lives in a two-dimensional
surface in the ($H, \, \dot{H}, \, \ddot{H}$) phase space. This surface is an
invariant manifold under the dynamics of eq. (\ref{order3}) which was obtained 
for quantum matter.
It is the initial conditions that distinguish the two cases: in the classical
case there is a precise relation between $H(t_0)$,$\dot{H}(t_0)$ and 
$\ddot{H}(t_0)$ which is not satisfied in the quantum case as can easily 
be seen, for example, for the vacuum or a coherent state.
The trajectories for a generic quantum state do not lie on this surface,
but they approach it asymptotically during evolution, however they will 
never exactly lie on it (see Fig. 1).

\begin{figure}
\centering
\includegraphics[width=3.0in]{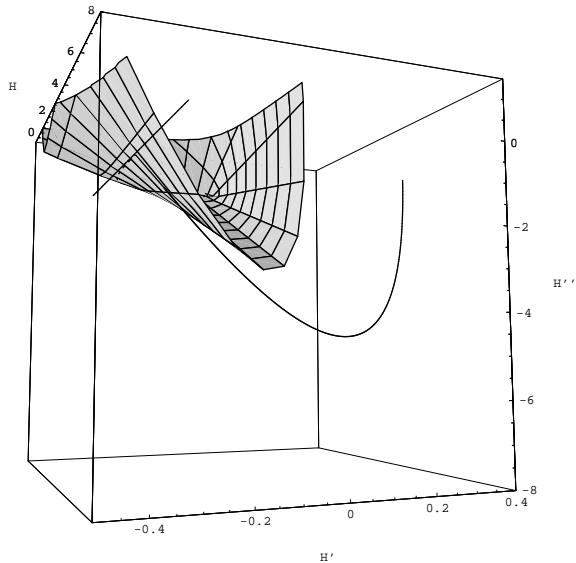}

\caption{In gray we have the invariant manifold corresponding to the 
classical dynamics (\ref{classic}). 
Also exhibited are a trajectory related to a highly non-classical 
state with initial conditions ($H=4$, $\dot{H}=0.4$, $\ddot{H}=0$)
and a line, given by ($\dot{H}=-1/3$, $\ddot{H}=0$) and not
intersecting the invariant surface, which is
approached by the system in the inflationary phase both for classical 
and quantum matter; after reaching it the trajectory will lie close to 
the invariant surface. Let us note that in the figures (1 and 2) we use
a prime to denote a time derivative.}

\end{figure}      

In order to understand how the inflationary regime can be reached one may 
study the set of initial conditions in the 
three dimensional ($H$,$\dot{H}$,$\ddot{H}$) phase space 
which lead to it.
Let us note that the only fixed point in the finite region for the vector 
field associated with
eq. (\ref{order3}) is given by $(H,\dot{H},\ddot{H})=(0,0,0)$. 
This point is asymptotically stable and the divergence of the vector 
field is negative everywhere for positive $H$, so that every trajectory will 
fall towards it. 

On the basis of the numerous analyses already performed on the classical 
system \cite{piran1,madsen,russi,amend} one expects
the value of $H$ to change during the eventual inflationary phase, decreasing
(on starting from initial conditions on the two-dimensional 
invariant manifold shown on Fig. 1)
from values generally much greater than one in the units we use (Planck mass).
This implies, on comparing the magnitude of the different terms of the 
eq. (\ref{order3}), that one can expect, from
the last two terms, unless $H$ is small,
\beq
\ddot{H} \approx 0 \quad , \quad 
\dot{H} \approx -\frac{1}{3} \,.
\label{phaseinfl}
\eeq
\noindent
Let us note that this is the same behaviour as the classical case, 
and is quite different from a true de Sitter evolution for which $H=const$.
We remark that this is 
the same behaviour as is also found by the authors of \cite{piran1}, who 
describe the curve corresponding to (\ref{phaseinfl}) by 
$\dot\phi_c \approx const$. 
Therefore in the three dimensional phase space, the trajectories 
starting from different points ( which are related to the initial conditions), 
for both the classical and quantum matter cases will,
if leading to an inflationary regime and after an eventually transient phase,
converge to a region around the curve characterized by 
(\ref{phaseinfl}) and a decreasing $H$.
This is what is shown and commented in Fig. 1.
Since one is also interested in the pressure/energy-density ratio, 
let us write the expression to be used later in the numerical simulations:

\beq
\frac{p}{\mu}=-1-\frac{2\dot{H}}{3H^2}.
\eeq
remembering that it is valid for any quantum state satisfying the 
Schr\"odinger equation.

The above approach for gravitation is general and does not require any
direct solution of the Schr\"odinger equation for matter.
However it is important to know the matter state, and for this reason we 
introduced the technique of the invariants.
On using them it is possible to construct coherent states 
to study the classical limit and estimate if a quantum state, 
satisfying the Sch\"odinger equation, defined by the Hamiltonian $\hat{H}_M$
will be modified by RHS fluctuations, possibly before the end of inflation. 
We shall attempt perturbative estimates in section 4.

\subsection{Matter states.}


For the coherent state one has 
a four dimensional dynamical system, 
which is given by eq. (\ref{rho}) for $\rho$ and:
\bea
\left( \frac{\dot{a}}{a}\right)^2 = 
\frac{8 \pi}{3} G ( \mu_{cl} + \mu_0 ) 
\label{vinc} \\
\dot{\Theta}=\frac{1}{a^3 \rho^2} \,
\eea   
We are interested in evolution from an initial state which is well
classified in terms of the Hamiltonian at the initial time ($t=t_0$),
that is a coherent state constructed from the particle annihilation 
and creation operators which factorize the Hamiltonian. 
This choice corresponds to requiring the invariant and the Hamiltonian
coincide at $t=t_0$ and the Bogoliubov coefficients will be 
$A(t_0)=0$ and $B(t_0)=1$, leading to the conditions:
\bea
\rho^2(t_0)&=&\frac{1}{m a^3(t_0)} \nonumber \\
\dot{\rho}(t_0)&=&0 \label{initphys}
\label{initcond}
\eea
Subsequently we shall consider a larger set of initial states not satisfying
the above.
It is remarkable that the above requirement relates 
the fluctuations of the scalar field given by eq. 
(\ref{indet}) to the scale factor of the universe.
Indeed at the quantum level the scale factor $a$
appears inextricably in the equations (already in the
energy density, see for example eq. (\ref{cohehami})), this is the
basic reason for which the semiclassical system is
three dimensional. From eq. (\ref{initcond}), corresponding to a possible
choice of initial conditions, it is clear that the smaller is the scale
factor the larger is $\rho$, which would actually be infinite if $a$ were
zero as is classically allowed.

Let us now examine the vacuum state.
This state is interesting even if it is obvious that 
it will be the first to be changed by the transitions generated 
by the RHS fluctuations of the matter equation (\ref{sch1}). 
The vacuum case corresponds to $\alpha=0$ ($\mu_{cl} = 0 $ in eq. 
(\ref{vinc})), so that $\Theta$ decouples and we may eliminate 
$a$ in eq. (\ref{rho}) obtaining an equation for $\rho$ and $H$. 
In analogy with eq. (\ref{passo1}), on scaling $\rho$ by $\sqrt{G\hbar}$, 
one also obtains:
\bea
\ddot{\rho}+3 H \dot{\rho}+\frac{\dot{\rho}^2-\frac{3}{2 \pi} H^2}{\rho}
+2 \rho = 0
\label{din2rho} \\
\dot{H} + 3 H^2 -2 \pi \rho^2 = 0 \,.
\label{din2h}
\eea  
\noindent
In this case we study the evolution of an initial state annihilated
by the invariant destruction operator which is not necessarily the Hamiltonian 
one, unless the conditions given in (\ref{initphys}) are imposed, and 
parametrize it in terms of the initial conditions ($\rho(t_0)$,
$\dot{\rho}(t_0)$ and $H(t_0)$) or ($H(t_0)$, $\dot{H}(t_0)$ and
$\ddot{H}(t_0)$). We observe that if the scalar field is in the n 
eigenstate of the invariant , the dynamical system is given by eq. 
(\ref{din2rho}) with the term $H^2$ divided by $(2n + 1)$ and by eq. 
(\ref{din2h}) with the term $\rho^2$ multiplied by $(2n + 1)$.

We may now consider a thermal (not pure) initial state which is also
characterized by zero mean values of the matter field and of its derivative. 
Such a state could arise 
within the context of our BO approach on introducing a complex, rather 
than real, time, thus generalizing our previous considerations. 
In order to describe such an initial state we consider a density 
matrix constructed
using as basis a given number of quanta of the Hamiltonian (at $t=t_0$).
If at time $t_0$ the conditions given in (\ref{initphys})
are satisfied, then the time evolution of
the basis will be given exactly by the eigenstates
of the quadratic invariant so that the density matrix will be
\beq
\hat{\rho}=(1-e^{-\beta m}) \sum_k e^{-\beta m k} |k,t \rangle \langle k,t |
\eeq
where $\beta=1/k_B T$ and of course during evolution it will no longer be 
a thermal state of the Hamiltonian.
The operator $\hat{\rho}$ satisfies the Liouville equation
so that also for this state one easily derives
the effective equation (\ref{order3}).
In fact using the known evolution equation for the $ \ \hat{\rho} \ $ operator 
we can compute, on taking traces, the mean value of all interesting operators.
In particular one has for the average matter Hamiltonian density:
\beq
Tr \left( \hat{\rho} \frac{{\cal H}_M}{a^3} \right) =
\mu_0 \left( 1+\frac{2}{e^{\beta m}-1} \right) \,
\label{thermalhami}
\eeq
and for the matter fluctuations
\beq
Tr (\hat{\rho} \ \hat{\phi}^2) =  
\frac{\hbar}{2} \rho^2 \bigl(  1+\frac{2}{e^{\beta m}-1}\bigr) \,.  
\eeq
These differ from the vacuum case by an extra factor which increases with
the temperature, as expected.
Thus, even for not so large values of $\rho$ we could still have large 
fluctuations of the matter field if the temperature is high.

\section{Numerical Simulations.}

In this section we exhibit the results of the numerical analyses of the 
previously described dynamical situations.

Let us first consider the basic effective equation (\ref{order3}), which is 
valid for every quantum system within the framework of the approximations 
made. 
We find that, starting from initial conditions very far from the 
``classical'' surface, the inflationary
regime coincides with the region described by eq. (\ref{phaseinfl}) and the 
trajectory will lie close to the two-dimensional invariant surface
previously mentioned. 
This is illustrated in Fig. 1.

In order to have a better idea of the different possible trajectories 
arising from differing initial conditions we show in Fig. 2 a projection 
of the trajectories on the $H$-$\dot{H}$ plane. 
From the picture it is evident that trajectories will lie,to within a very 
good approximation, for some time on the curve given by (\ref{phaseinfl}).
Therefore it is evident that we have a large set of initial conditions 
which will lead to an inflationary stage.
One also sees that the number of e-folds reached at the end of the 
inflationary stage depends on the value of $H$ at the moment when each
trajectory approaches this curve. 
In fact it can be seen that during the 
inflationary stage starting at $t=t_i$ one has 
\beq 
\frac{a(t)}{a(t_i)} \approx \exp \left( H(t_i) \,t + \dot{H}(t_i) 
\frac{t^2}{2} \right)\,
\eeq
and thus the total number of e-folds is
\beq
N = \int^{t_f}_{t_i} H(t) dt \approx -\frac{H^2(t_i)}{2 \dot{H}(t_i)} \,
\label{efoldsH}
\eeq
where $t_f$ denotes the time at which inflation stops and we have considered 
$H(t_f) \ll H(t_i)$.

\begin{figure}
\centering
\includegraphics[width=3.0in]{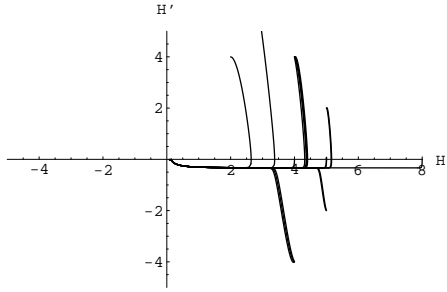}\\

\caption{Trajectories with different initial conditions projected on the
$H$-$\dot{H}$ plane. The bold lines reflect a superpositions of curves
having quite different $\ddot{H}$ for some time. It is evident that all the 
curves reach the common region described by eq. (\ref{phaseinfl}).}
\end{figure}      

In order to compare our results with the already known behaviour of classical
models \cite{piran1,madsen,russi,amend}
we shall start from a system described by a coherent state which will be
squeezed during time evolution.
The results of the analysis are easy to understand
from the structure of equation (\ref{vinc}) on noting that $H$ (or $a$) is 
completely governed by the matter Hamiltonian density which has two 
contributions, one from the vacuum and another which is exactly
that expected from classical dynamics.

Therefore if we are in a coherent state characterized by an average number 
of quanta $|\alpha|^2 \gg 1$, the vacuum contribution will be negligible and 
the resulting evolution will be close to the
one given by the classical homogeneous matter field.
For such a case we show in Fig. 3 the e-fold number, 
the Hubble parameter, the mean field fluctuation, the pressure/energy-density 
ratio and the average field $\bra{\alpha} \hat{\phi} \ket{\alpha}$, which 
corresponds to the classical quantity $\phi_c$, resulting from the dynamics. 
We start from a condition, known to lead to inflation, with the average 
value of the field starting the slow rollover phase; 
more precisely, working in units of the Planck mass, we have taken for the
inverse Compton wavelength $m=10^{-6}$, imposed initial condition 
(\ref{initphys}) and $|\alpha|=2.334 \ 10^3$, $\beta =0$, 
$\rho(t_0)=10^{-3}$, $\dot{\rho}(t_0)=0$ so that $\phi(t_0)=3.3$, 
$\dot{\phi}(t_0)=0$.
The resulting initial conditions for
equation (\ref{order3}) are $H=6.7539 \,,\, \dot{H}=-6.283 \ 10^{-6}$ and 
$\ddot{H}=6.878 \ 10^{-15}$ (all these numerical values 
are for the rescaled dimensionless quantities and time scaled by $m^{-1}$).
After a short transient, a linear decrease of $H$ and $\rho$ until the 
inflation period ends (with an e-fold number of $70$) is evident. 
The behaviour of the ratio between the pressure and the
energy density remains, as expected, close to a value of $-1$ for most
of the inflationary period and later starts to oscillate.

From the point of view of the matter state, it is interesting to note that the
role played in the classical case by the value of the field is now replaced by
$\rho$, which is related to the quantum fluctuations of the homogeneous field.
During inflation $\rho$ decreases linearly and after this ``dissipation'' 
it will start to oscillate above zero.
We also note that the average value of the field behaves as the corresponding
classical quantity, decreasing during inflation and oscillating after it
(in the figure we use a logarithmic scale which covers only positive values).

\begin{figure}
\centering
\raisebox{4cm}{$\ln{\frac{a(t)}{a(t_0)}}$} \hspace {-0.2cm} 
\includegraphics[width=2.9in]{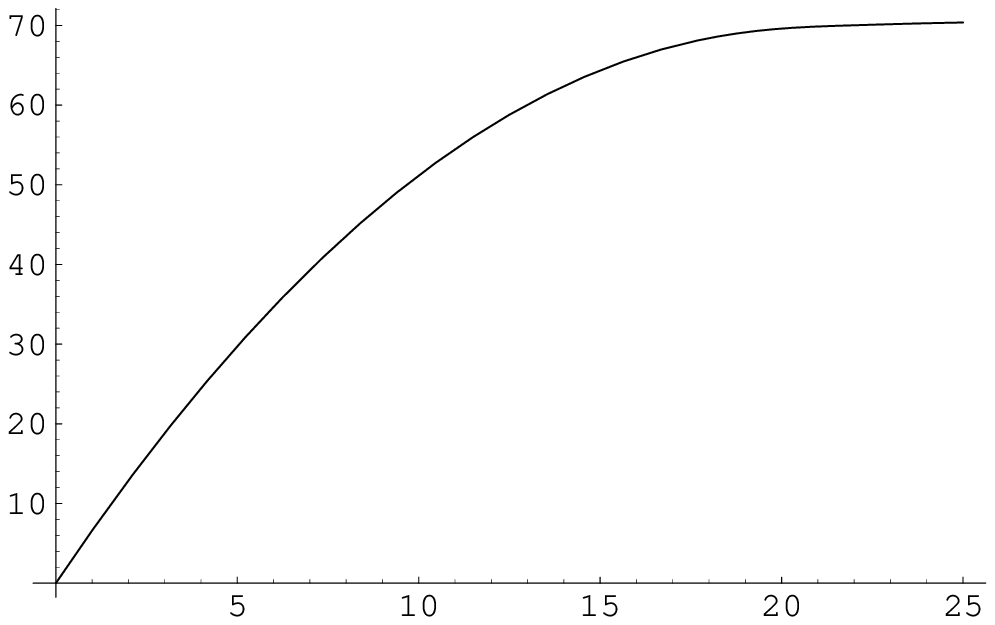}
\hspace{-0.2in}
\raisebox{4cm}{$H$} \hspace {-0.3cm} 
\includegraphics[width=2.9in]{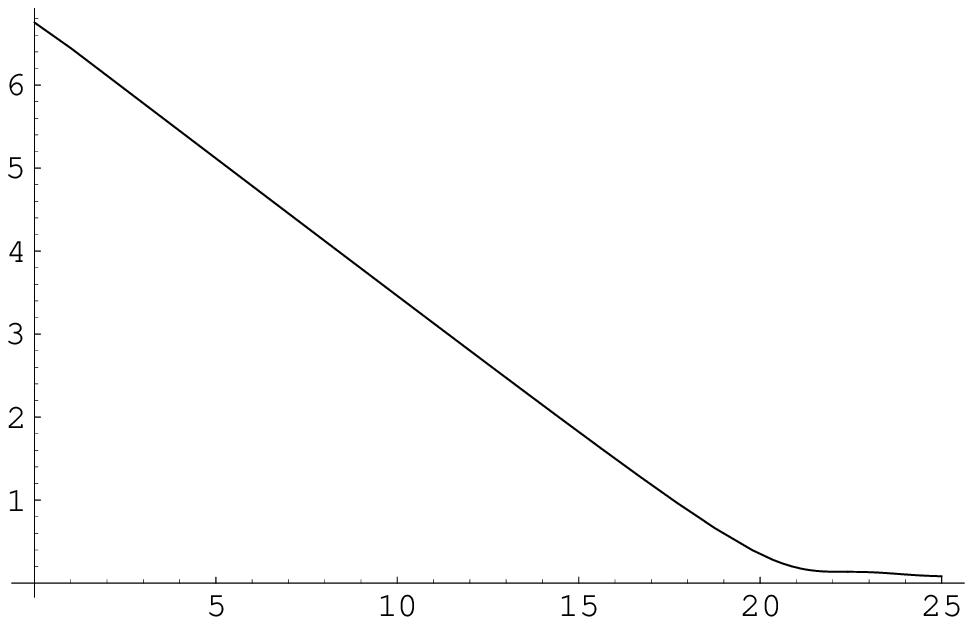}
\raisebox{0.5cm}{\hspace{1cm} $t$ \hspace{7cm} $t$}\\

\raisebox{4cm}{$\rho$} \hspace {-0.2cm} 
\includegraphics[width=2.9in]{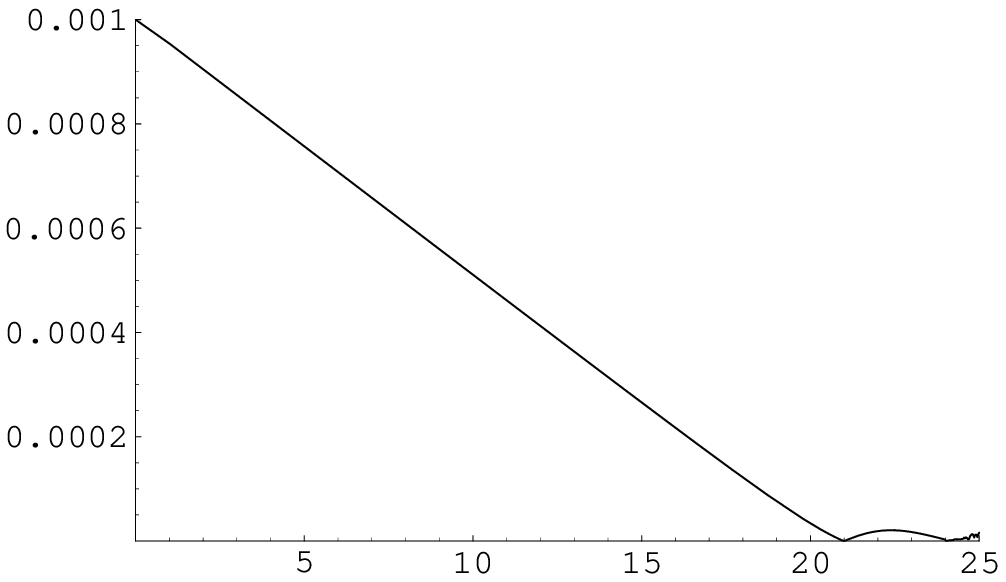}
\raisebox{4cm}{$\frac{p}{\mu}$} \hspace {-0.4cm} 
\includegraphics[width=2.9in]{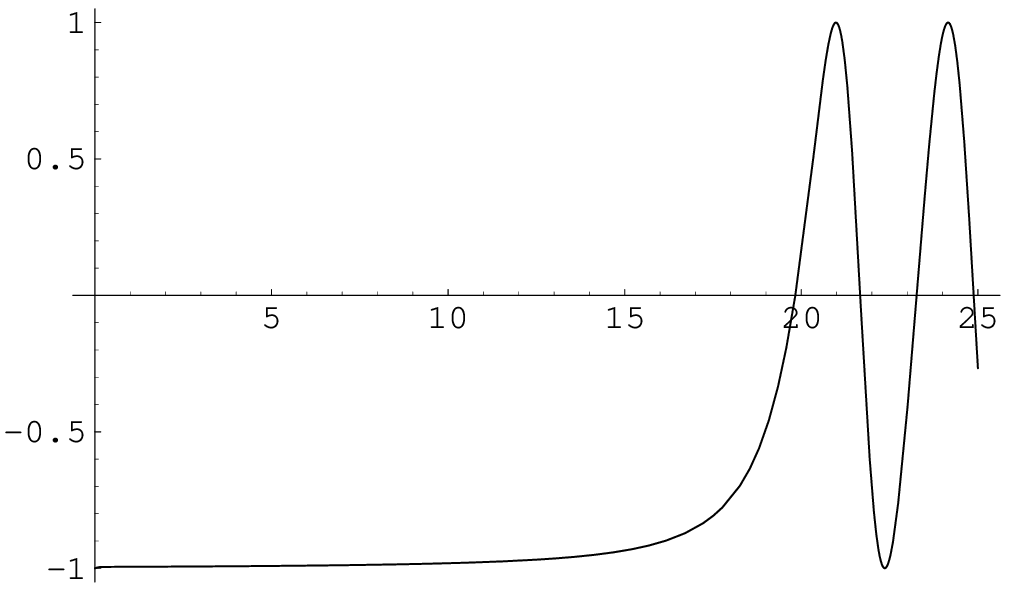}
\raisebox{2cm}{\hspace{-.2cm} $t$}\\
\raisebox{0cm}{\hspace{-6.5cm} $t$}

\raisebox{4cm}{$\phi_{c}$} \hspace {-0.5cm} 
\includegraphics[width=2.9in]{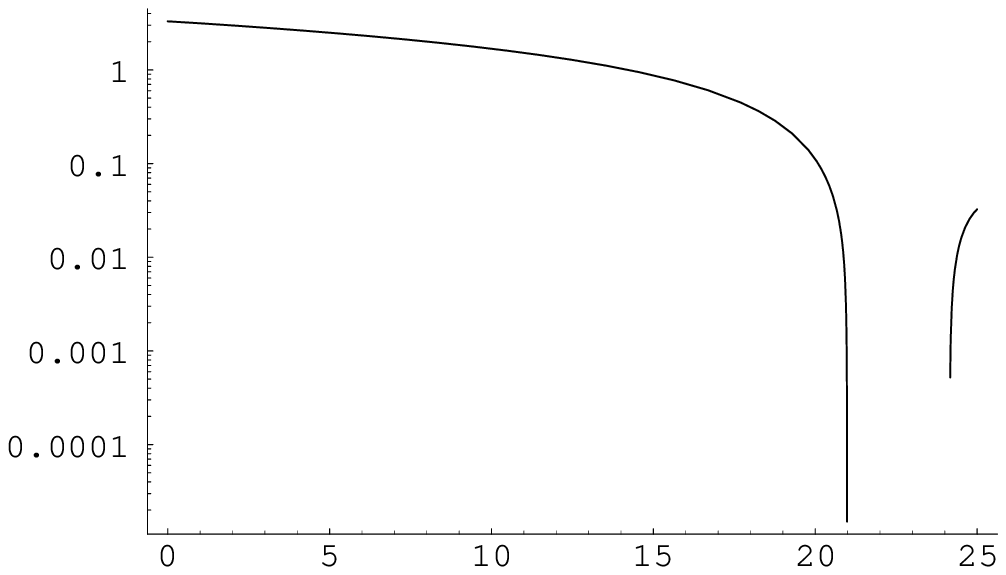}\\
\raisebox{1cm}{\hspace{1cm} $t$}

\caption{For matter in a coherent state we show the time behaviour
of the e-fold number, the Hubble parameter $H$, the mean value $\rho$ for the 
fluctuations of the field, the pressure/energy-density ratio and of the 
average value of the field $\phi_c=\langle \phi \rangle_\alpha$.}                                                         
\end{figure}

Let us consider the other extreme case: that of the  
invariant vacuum (which eventually can also be the Hamiltonian vacuum 
at $t=t_0$). 
For this state the average field and time derivative of the field are zero.
On neglecting the transitions induced by the RHS of the matter
equation, it is clear that for high enough fluctuations 
(large values of $\rho$)
the system could undergo an inflationary phase. In such a case 
the behaviour of $a$, $H$, $\rho$, $p/\mu$ is the same as that of the coherent 
case plotted in Fig. 3.
However the presence of a non negligible term on RHS implies that 
the Schr\"odinger equation will be a rough approximation to the real matter 
equation and we are then unable to predict the evolution of the system.

Since the vacuum is an extreme case we feel that
it is interesting to try to further analyze such a situation.
Let us then  estimate perturbatively the time necessary for the corrections
to produce a large transition rate such that it would be meaningless
to consider the normal (Schr\"odinger) evolution for the vacuum.
After such a time we shall no longer be able to make any prediction
and if the system was, or was about to enter, in an inflationary stage it 
may or may not thus continue.
We rewrite the full matter equation (with the RHS) as
\beq
\Bigl( i \hbar \frac{\partial}{\partial t} -  \hat{H}_M \Bigr) \ket{\chi}_s
= - e^{i \varphi} \hat{F}_{\chi} \ket{\chi}
\label{fullmat}
\eeq
where $\varphi$ and $\hat{F}_{\chi}$ are defined by comparison with eqs.
(\ref{chisch}) and (\ref{sch1}).
The full higher order non-linear equation is clearly too complicated to solve.
Hence we estimate the transitions to other states which are solutions of the
unperturbed Schr\"odinger equation, as is customarily done in perturbation
theory.
This means we expand $\ket{\chi}_s=\sum_n c_n(t) \ket{n}_s$ and
evaluate the contributions of the perturbation to lowest order \cite{kim}, 
thus we consider the operator $\hat{F}_0$ which depends on the $\ket{0}$
state.
One then obtains for the transition rate:
\beq
  - i \hbar \frac{\partial c_n}{\partial t}=e^{i n \Theta} 
\bra{n}\hat{F}_0 \ket{0}
\label{breakrate}
\eeq
Further we must choose the initial values for $\rho$, $\dot{\rho}$ and $a$.
The following relations, obtained from eqs. (\ref{cohehami}) and (\ref{vinc}) 
are useful for this purpose:
\beq
H=\sqrt{\frac{2 \pi G \hbar}{3}}\sqrt{\rho^2-\frac{\dot{H}}{2\pi G \hbar}}, 
\quad  
\dot{H}=-2 \pi G \hbar \Bigl(\dot{\rho}^2+\frac{1}{m^2 a^6 \rho^2} \Bigr), 
\quad
\ddot{H}=-6 H \dot{H} + 4 \pi G \hbar \rho \dot{\rho}
\label{phasevac}
\eeq

We first consider the case for which the state is initially
in an inflationary regime so that the relations
in (\ref{phaseinfl}) are satisfied. On requiring a certain number of e-folds
we find the value of $H$ needed and hence 
the necessary $\rho$ and $\dot{\rho}$ and $a$.
For example values $H>1$ will lead to $\rho \approx 0.7 H$ and 
$\dot{\rho} \approx -0.23$ in the chosen units.
\begin{figure}
\centering
\raisebox{4cm}{${\cal R}e \, \dot{c}_4$} \hspace {-0.2cm} 
\includegraphics[width=3.0in]{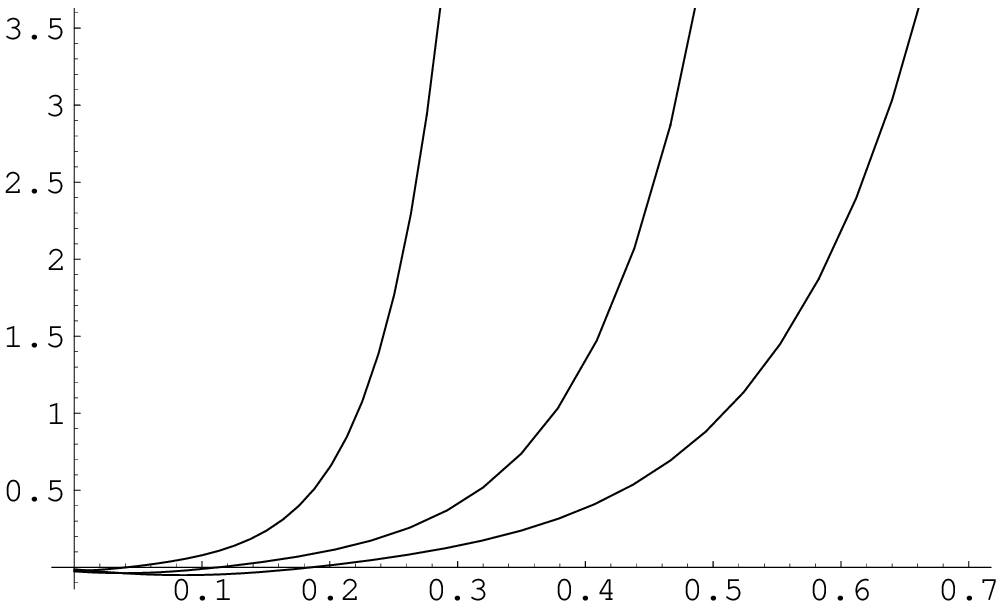}\\
\raisebox{.5cm}{\hspace{1.5cm} $t$}

\caption{For an inflationary region with different values for 
$\rho$ (for $H$) which would lead, in absence of RHS correction, to
an inflation with e-fold number of $70$, $25$, $15$ , the corresponding
(from left to right) time behaviour of the transition rate eq. 
(\ref{breakrate}) is shown.
For the above the transition rate becomes significant and possibly modifies
the inflationary regime after $1\%$, $4\%$ and $6\%$ respectively of the 
inflationary period in the absence of RHS fluctuation corrections.}
\end{figure}
On using these values we have computed the quantity in eq. (\ref{breakrate}).
In particular in Fig. 4 we show the real part of $\dot{c}_4$ for three
different cases leading to an e-fold of $70$, $25$ and $15$ if the
corrections were not present
(the right hand side of Eq. (\ref{breakrate}) has comparable real and 
imaginary parts, for 
this reason the contribution to the real part of $\dot{c}_4$, plotted 
in Fig. 4, due to a possible rapidly oscillating phase of $c_4$ is 
excluded).
We note that for smaller initial $H$ (which would lead to less expansion)
the transition rate becomes significant at a later time.
If, on the other hand, we consider an initial vacuum satisfying the 
conditions of eq. (\ref{initphys}) (as yet we are not in an inflationary
regime),
the transition rate is non-negligible from the beginning and as a consequence
we simply cannot use the Schr\"odinger equation to evolve it
(analogous considerations hold for $c_2$).
 
Little can be added on the evolution from a thermal state,
on neglecting the RHS, since it is similar the vacuum case.
The basic difference is that the thermal factor in eq. (\ref{thermalhami})
will increase, depending on the temperature, the energy density.


Let us end this section by adding a brief comment about {\em reheating}.
It is known that an inflationary stage during which the scale 
factor is accelerated should be associated with a massive entropy 
production in order to solve the problems of the standard 
cosmology \cite{libri}. However, also for states of the matter field
with a vanishing mean value persisting till the end of an inflationary phase,
$\rho$ would oscillate with the same period the classical field 
would have had, but its amplitude would decrease in a more complicated way 
(see, for instance, eq. (\ref{din2rho}) for the vacuum case). 
The similarity of the behaviour of scalar field fluctuations for classical 
and vacuum states has already been noted \cite{desitter} in an 
exact de Sitter space-time for long wavelength modes. For these reasons 
in a model where
a second field $\chi$ is coupled through a $\phi^2 \chi^2$ term to 
the inflaton, the vanishing value of the scalar field does not seem to 
prevent the possibility of a reheating process after inflation.  
In any case we realize that the inclusion of all the other modes is 
necessary to tackle this program \cite{suen}.     
\section{Conclusions}

In this paper we have studied a massive homogeneous minimally coupled 
scalar field in a flat RW metric. After having quantized the 
system we recovered, through  a BO decomposition, the semiclassical evolution 
for gravity and a Schr\"odinger equation for matter, always on neglecting 
RHS fluctuation corrections. 
Such a treatment is however an approximation which could 
break down for a matter quantum state because of the growth of the corrections.
Since we do not know how to predict exactly the dynamical consequences 
of the RHS fluctuations, we evaluated them perturbatively and  
we studied in detail the properties of the
dynamical system and the time scale at which the corrections are no 
longer negligible. 

In particular we found:

(1) The dynamical semiclassical gravity system for a generic quantum 
state of the scalar field is described by a non-linear third order 
differential equation for the Hubble parameter. 
We have seen that the classical 
system lives in an invariant surface of the three dimensional phase 
space and that an inflationary regime (with $\dot H \simeq - \frac{1}{3}$) is 
experienced for a generic quantum state of the scalar field.

(2) The time scale at which the corrections 
are no longer negligile depends on the "classicality" of the quantum state 
for the scalar field. In the classical limit ($|\alpha|^2 
\rightarrow + \infty$) these corrections are suppressed while for a generic 
quantum state they are no longer negligible before inflation ends. 
For the vacuum state the corrections could become important,
in the least favorable case (for example by choosing particular initial 
conditions), at the same time scale as that for which inflation starts. 

\noindent
In any case we feel that from this analysis it emerges that inflation 
can arise from a quantum mechanical initial matter state, or at least 
a coherent one, and not just a classical state. Such a situation was 
envisaged in the earlier pioneering works \cite{brout, private}.

Let us conclude by again noting that the main problems encountered in
considering inflation from a non-classical state of the scalar field are
related to the difficulties in solving the full matter equation
(\ref{fullmat}).
One would like to at least evaluate the effects of the RHS fluctuations 
non-perturbatively and see how an inflationary regime evolves from a generic
quantum matter state.
We intend to investigate this and other points in more detail, as well as 
applications of the above method to other physical problems 
\cite{faccioli}.  
\section{Acknowledgements}
We wish to thank Roberto Balbinot and Robert Brout for 
useful discussions.
One of us (F. F.) would like to thank Marco Bruni for helpful comments.

\section{Appendix}

We here compute the time derivatives of the $|n\rangle_b$ and of the coherent 
states for the destruction operator $b$ satisfying
\beq
\frac{\partial \hat b}{\partial t} = \frac{i}{\hbar} [\hat b, \hat H_M] 
- i \dot \Theta \hat b \,.
\label{bevol}
\eeq
\noindent
The coherent state $\ket{\alpha}_b$ is defined by applying the operator
$\hat D$
\beq
\hat D = e^{\alpha \hat b^{\dagger} - \alpha^* \hat b} = 
e^{-\frac{|\alpha|^2}{2}}\, e^{\alpha \hat b^{\dagger}} \, 
e^{- \alpha^* \hat b} 
\eeq
to the vacuum state $\ket{0}_b$, for which $\hat b \, \ket{0}_b =0$.
Therefore 
\beq
\frac{\partial \ket{\alpha}_b}{\partial t} = \frac{\partial \hat D}
{\partial t} \, \ket{0}_b + \hat D \, \frac{\partial \ket{0}_b}{\partial t} 
= e^{-\frac{|\alpha|^2}{2}}\, 
\sum\limits_{n=0}^\infty\, {\alpha^n\over\sqrt{n!}} 
\frac{\partial}{\partial t} {| n \rangle}_b
\label{dercoh0}
\eeq
with
\beq
\frac{\partial}{\partial t} {| n \rangle}_b=\left[
\frac{\partial}{\partial t} \hat b^{\dagger \, n} + \hat b^{\dagger \, n} 
\frac{\partial}{\partial t} \right] \ket{0}_b  
\,
\label{dern0}
\eeq
where, on recalling the equations satisfied by $\chi$, we have 
for the second term in (\ref{dern0})
\bea
\frac{\partial \ket{0}_b}{\partial t} &=& \left( \, _b\bra{0} \frac{\partial}
{\partial t} + \frac{i}{\hbar} \hat H_M \ket{0}_b \right) \ket{0}_b 
- \frac{i}{\hbar} \hat H _M \ket{0}_b = \nonumber \\
&=& _b\bra{0} \frac{\partial}{\partial t} \ket{0}_b
- i m B A^* \hat{b}^{\dagger\,2} \ket{0}_b \,.
\eea

On using 
\beq
\frac{\partial \hat b^{\dagger}}{\partial t} 
= \frac{i}{\hbar} [\hat b^{\dagger}, \hat H_M]
+ i \dot \Theta \hat b^{\dagger} \,.
\label{bconevol} \,
\eeq 
and 
\beq
[\hat b^{\dagger}, \hat H_M] = \hbar m [- (1 + 2 |A|^2)\, b^{\dagger} - 
2 A B^* \hat b] \,
\eeq
we have 
\beq
\frac{\partial}{\partial t} \hat b^{\dagger \, n}=i\bigl[\dot \Theta - m
(1+2 |A|^2)\bigr] n b^{\dagger \, n} -2 i m B^* A 
\bigl[ \frac{n (n-1)}{2} \hat b^{\dagger \, n-2} + \hat b^{\dagger \, n-1} 
\hat b \bigr]
\eeq
One finally obtains:
\bea
\frac{\partial}{\partial t} \ket{n}_b&=&\bigl\{ i n 
\bigl[ \dot \Theta -m (2 |A|^2+1) \bigr] +
_b\bra{0} \frac{\partial}{\partial t} \ket{0}_b \bigr\} \ket{n}_b \nonumber \\
&& -i m B^* A \sqrt{n(n-1)} \ket{n-2}_b
   -i m B A^* \sqrt{(n+1)(n+2)} \ket{n+2}_b
\label{dern}
\eea
and
\beq
\frac{\partial}{\partial t} \ket{\alpha}_b = \left\{ -i m B A^*
\hat b^{\dagger\,2}
+ \alpha \left[ i\dot\Theta - i m \left(1 + 2|A|^2\right)
\right] \hat b^{\dagger} +
\left[ _b\bra{0} \frac{\partial}{\partial t} \ket{0}_b -
i m A B^* \alpha^2 \right]
\right\} \ket{\alpha}_b \,.
\label{dercoh}
\eeq
where
\beq
_b\bra{0} \frac{\partial}{\partial t} \ket{0}_b=
-i \dot{\varphi}_0 -i m (|A|^2+\frac{1}{2})
\eeq


\begin{thebibliography}{3}

\bibitem{brout}
R. Brout, F. Englert and E. Gunzig, {\em Gen. Rel. Grav.} {\bf 10} (1979) 1;
A. A. Starobinsky, {\em Phys. Lett.} B {\bf 91} (1980) 99
%
\bibitem{guth}
A. H. Guth, {\em Phys. Rev.} D {\bf 23} (1981) 347
%
\bibitem{linde1}
A. D. Linde, {\em Phys. Lett.} B {\bf 129} (1983) 177
%
\bibitem{new}
A. Albrecht and P. J. Steinhardt, {\em Phys. Rev. Lett.} {\bf 48} (1982) 
1220; A. D. Linde, {\em Phys. Lett.} B {\bf 108} (1982) 389
%
\bibitem{linde2}
A. D. Linde, {\em Phys. Lett.} B {\bf 162} (1985) 281
%
\bibitem{initial}
D. S. Goldwirth and T. Piran, {\em Phys. Rept.} {\bf 214}, (1992) 223
%
\bibitem{rev}
K. Olive, {\em Phys. Rept.} {\bf 190}, (1990) 307
%
\bibitem{brand}
V. F. Mukhanov, H. A. Feldman and R. H. Brandenberger, {\em Phys. Rept.} 
{\bf 215}, (1992) 203 
%
\bibitem{cv}
R. Casadio and G. Venturi, {\em Class. Quantum Grav.} {\bf 13}, (1996) 2715
%
\bibitem{lewis} H. R. Lewis and B. Riesenfeld, {\em J. Math. Phys.} {\bf 10}
                (1969) 1458
%
\bibitem{qc}
J. B. Hartle and S. W. Hawking, {\em Phys. Rev.} D {\bf 28}, (1983) 2960; 
A. Vilenkin, {\em Phys. Rev.} D {\bf 33} (1986) 3560 
%
\bibitem{barvinsky}
A. O. Barvinsky, A. Yu Kamenschik and I. V. Mishakov, {\em Nucl. Phys.} 
B {\bf 491}, (1997) 387
%
\bibitem{klauder} J. R. Klauder and B.-S. Skagerstam, {\em 
Coherent States - Applications in Phisics and Mathematical Physics}, 
World Scientific (1985)
%
\bibitem{coh} J. G. Hartley and J. R. Ray, {\em Phys. Rev.} D
{\bf 25} (1982) 382; J. A. Pedrosa, {\em Phys. Rev.} D {\bf 36} (1987) 1279
%
\bibitem{piran1} T. Piran and R. M. Williams, 
{\em Phys. Lett.} B {\bf 163} (1985) 331; 
T. Piran, {\em Phys. Lett.} B {\bf 181} (1986) 238
%
\bibitem{brout2}
R. Brout, {\em Found. Phys.} {\bf 17} (1987) 603;
%
\bibitem{bv}
R. Brout and G. Venturi, {\em Phys. Rev.} D {\bf 39} 2436 (1989)
%
\bibitem{dewitt} 
J. A. Wheeler, in {\em Relativity, Groups and Topology}, edited by 
C. M. DeWitt and J. A. Wheeler, Benjamin, New York (1968); 
B. S. DeWitt, {\em Phys. Rev.} {\bf 160}, 113 (1967)
%
\bibitem{banks}
T. Banks, {\em Nucl. Phys.} B {\bf 249} (1985) 332;
%
\bibitem{cinesi} X.-C. Gao, J.-B. Xu and T.-Z. Quian {\em Phys. Rev.} A 
{\bf 44} (1991) 7016
%
\bibitem{berry} A. Messiah, {\em Quantum Mechanics}, Vol. 2, Interscience, 
New York (1961); M. V. Berry, {\em Proc. Roy. Soc.} London Ser. {\bf A392}, 
(1984) 45
%
\bibitem{bertoni} C. Bertoni, F. Finelli and G. Venturi, {\em Class.
Quantum Grav.} {\bf 13} (1996) 2375
%
\bibitem{madsen} M.S. Madsen and P. Coles, {\em Nucl. Phys.} B 
{\bf 298} (1988) 701
%
\bibitem{russi} V. P. Belinsky, L. P. Grishchuk, I. M. 
Khalatnikov and Ya. B. Zeldovich, {\em Phys. Lett.} B {\bf 155}
(1985) 232
%
\bibitem{amend} L. Amendola, M. Litterio and F. Occhionero, 
{\em Int. J. Mod. Phys.} A {\bf 5} (1990) 3861
\bibitem{cooper} The difference between the exact full quantum evolution 
of a system and its approximation is well known. For instance this aspect 
has been recently analyzed for a different coupled system by 
F. Cooper, J. Dawson, S. Habib and R. D. Ryne, preprint quant-ph/9610013 
%
\bibitem{kim} A related attempt to evaluate fluctuations has been done by 
S. P. Kim, {\em Phys. Rev.} D {\bf 55} (1997) 7511
%
%
%
\bibitem{libri} E. W. Kolb and M. S. Turner, {\em The Early Universe}, (1990) 
Addison-Wesley; 
A. Linde, {\em Particle Physics and Inflationary Cosmology}, (1990) Harwood 
Academic Publishers 
%
\bibitem{desitter} See C. Bertoni, F. Finelli and G. Venturi, 
{\em Phys. Lett.} A {\bf 237}, (1998) 331 and references therein.
%
\bibitem{suen} W.-M. Suen and P. R. Anderson, {\em Phys. Rev.} D 
{\bf 35}, (1987) 2940 
%
\bibitem{private} After the manuscript was completed we 
received the following private communication from R. Brout:

\noindent
``The work of R. Brout, F. Englert and E. Gunzig in \cite{brout}, as 
well as the subsequent self-consistently driven approach of 
R. Brout, F. Englert and P. Spindel ({\em Phys. Rev. Lett.} {\bf 43} 
(1979) 417) did not give full justice to 
the infra-red sector of the fluctuations of the scalar field. 
Among other things, rescaling all the fields by the cosmological 
scale factor leads to surface terms which probably play a very 
important r\^{o}le in the infra-red and hence their neglect 
results in a distortion of the true physics. The analysis of the 
fluctuations of the homogeneous mode in the present paper is a case 
in point. The physics is far richer than envisioned in our original 
work. ``
%
\bibitem{faccioli} L. Faccioli, F. Finelli, G.P. Vacca and G. Venturi,
{\em Phys. Rev. Lett.} {\bf 81}, (1998) 240


\end{thebibliography}
\end{document}